\documentclass[aps,prl,reprint,longbibliography]{revtex4-2}

\usepackage{graphicx}   
\usepackage{setspace}
\usepackage{dcolumn}    
\usepackage{bm}         
\usepackage{amsmath}    
\usepackage{amssymb}    
\usepackage{mathtools}

\usepackage{amsfonts}
\usepackage{xcolor}

\newcommand{\ket}[1]{{\left| #1 \right>}}

\newcommand{\expec}[3]{{\left< #1 \middle\vert #2 \middle\vert #3 \right>}}
\newcommand{\bracket}[1]{{\left< #1 \right>}}

\begin{document}

\title{SU($N$) Quantum Spin Model with Weak and Strong First-Order Néel to Valence-Bond Solid Transitions}

\author{Ryan Flynn}
\email{rflynn22@bu.edu}
\affiliation{Department of Physics, Boston University, 590 Commonwealth Avenue, Boston, Massachusetts 02215, USA}

\author{Anders W. Sandvik}
\email{sandvik@bu.edu}
\affiliation{Department of Physics, Boston University, 590 Commonwealth Avenue, Boston, Massachusetts 02215, USA}

\date{\today}

\begin{abstract}
We introduce an SU($N$) symmetric two-dimensional quantum spin model, the $X$-$Q$ model, which hosts a ground state transition between Néel antiferromagnetic and spontaneously dimerized states. The $Q$ terms are products of two adjacent singlet projectors on nearest-neighbor sites, as in the often studied $J$-$Q$ model (where $J$ is the Heisenberg exchange), while the $X$ terms are products of two permutation operators on second-neighbor sites. Quantum Monte Carlo simulations reveal close proximity to a deconfined quantum critical point for $N=2$, as in the $J$-$Q$ model. However, for $N>2$ the transition becomes strongly first-order, contrary to conventional expectations that increasing $N$ should weaken discontinuities. We attribute this behavior to the inability of the $X$ term to induce U(1) fluctuations of the dimer pattern, while those from the $Q$ term are suppressed by $1/N$. These results provide insights into the interactions that support deconfined criticality. 
\end{abstract}

\maketitle

Two-dimensional (2D) quantum antiferromagnets on the square lattice provide a central platform for the study of strongly correlated quantum matter \cite{Chakravarty_PRB_1989,Manousakis_RevModPhys_1991,Auerbach1994,Sachdev_WS_1995}. Their ground states are typically characterized by Néel antiferromagnetic (AFM) order, and understanding how this order is destabilized has been a long-standing problem \cite{Inui_PRB_1988,Anderson_MatResBul_1973,Anderson_Science_1987,Chubukov_PRB_1994,Lee_RevModPhys_2006,Sachdev2011,Scalapino_RevModPhys_2012} since the discovery of high-temperature superconductivity emerging in the cuprates upon doping \cite{Bednorz_ZPB_1986}. Another case, the transition from an AFM state to a spontaneously dimerized ground state (the valence-bond solid, VBS) \cite{Haldane_PRL_1988,Dagotto_PRL_1989,Read_PRL_1989,Read_PRB_1990,Harada_PRL_2003,Sachdev_NatPhys_2008,Beach_PRB_2009}, is of fundamental interest in its own right and also has potential intersections with the still unresolved high-$T_c$ problem \cite{Kaul_PRB_2007}. Following suggestive numerical findings \cite{Assaad_PRB_1997,Sandvik_PRL_2002,Motrunich_PRB_2004}, the theory of deconfined quantum criticality (DQC) \cite{Senthil_Science_2004,Senthil_PRB_2004,Levin_PRB_2004,Senthil_PRB_2006} posits that such a transition can be continuous, involving fractionalized excitations, an emergent gauge field, and enhanced symmetries. Such transitions lie beyond the conventional Landau–Ginzburg–Wilson paradigm, where a transition between ordered states breaking unrelated symmetries is generically of first-order.  

Early quantum Monte Carlo (QMC) results for the 2D $J$-$Q$ spin models hosting the AFM--VBS transition confirmed the key prediction of emergent U(1) symmetry of the microscopically $\mathbb{Z}_4$ symmetric VBS order parameter near criticality \cite{Sandvik_PRL_2007,Jiang_JStatMech_2008,Melko_PRL_2008,Lou_PRB_2009}, reflecting ``dangerous irrelevance'' of the lattice anisotropy; a well-known phenomenon in classical clock models \cite{Lou_PRL_2007,Okubo_PRB_2015,Shao_PRL_2020}. Subsequent studies of SU($N$) models \cite{Lou_PRB_2009,Kaul_PRB_2011,Kaul_PRL_2012,Block_PRL_2013} found quantitative agreement with large-$N$ field-theoretic predictions that the AFM--VBS transition is described by the non-compact CP$^{N-1}$ gauge field theory \cite{Senthil_PRB_2004,Dyer_JHEP_2015}.

\begin{figure*}[t] 
\includegraphics[width=176mm]{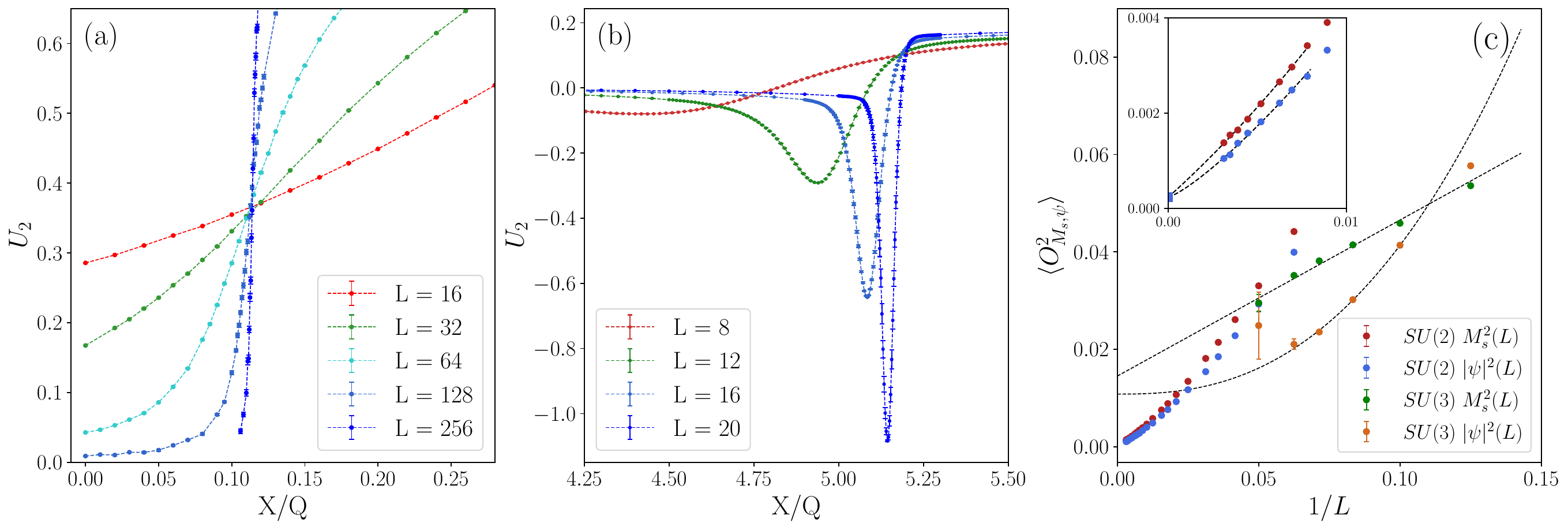}
\caption{Binder cumulant for (a) SU(2) systems with $L=16\text{–}256$ and (b) SU(3) systems with $L=8\text{–}20$. In (a) the fitted curves cross at the near-critical point $(X/Q)_c = 0.1152(2)$ while in (b) the negative peak flows to a clearly first-order transition at $(X/Q)_c = 5.223(2)$. (c) Extrapolations of the order parameters $M_s^2$ and $|\psi|^2$ at the transition point [defined by cumulant crossings for SU(2) and cumulant minima for SU(3)], with the non-zero $L \to \infty$ values [very small in the SU(2) case] demonstrating phase coexistence in both models. Polynomial fits were used for all order parameters, except $|\psi|^2$ for SU($3$), where the manifestly discrete symmetry breaking implies exponential convergence. The inset in (c) zooms in on the large-$L$ SU($2$) data, with extrapolated order for $L\rightarrow\infty$ indicating phase coexistence (error bars are smaller than the points).}
  \label{fig:Binder}
\end{figure*}

Despite this convergence of results, there is continued disagreement on the nature of the transition for small $N$. The SU(2) lattice models do not exhibit clean criticality, instead showing finite-size effects that have been interpreted either as weak first-order transitions \cite{Kuklov_PRL_2008,Jiang_JStatMech_2008,Harada_PRB_2013,Chen_PRL_2013,Nahum_PRX_2015,Zhao_PRL_2020,DEmidio_SciPost_2023} or anomalous corrections inherent to DQC \cite{Sandvik_PRL_2010,Nahum_PRL_2015,Shao_Science_2016}. In one scenario, the inability to reach the critical point exactly is a fundamental aspect of SU(2) DQC---its description by a complex conformal field theory (CFT) with an unphysical fixed point \cite{Ma_PRB_2020,Nahum_PRB_2020,Zhou_PRX_2024,Zou_arxiv_2025}. A competing view is that DQC arises from a multicritical point at the tip of the gapless spin liquid that has been observed in numerics  between AFM and VBS phases \cite{Gong_PRL_2014,Morita_JPhysSoc_2015,Wang_PRL_2018,Ferrari_PRB_2020,Nomura_PRX_2021,Wang_ChinPhysLett_2022,Liu_PRX_2022,Liu_PRB_2024,Viteritti_PRB_2025} and for which various field theories have been proposed \cite{Feuerpfeil2026,Shackleton_PRB_2021}. These spin liquids are out of reach of sign-problem-free simulations, and the multicritical point may therefore also be out of reach of QMC studies. The most recent large-scale QMC simulations of the $J$-$Q$ model \cite{Takahashi_2024} nevertheless demonstrate very close proximity to a critical point with emergent SO(5) symmetry \cite{Nahum_PRL_2015} and good agreement with exponents from related CFT calculations
\cite{Chester_PRL_2024,Li_arxiv_2026}.

These developments and the persistent controversy regarding $N=2$ (and other small $N$) \cite{Zhao_PRL_2022,DEmidio_PRL_2024,Zhou_PRX_2024,Song_ADV_2025} motivate the exploration of other lattice models that could potentially be tuned even closer to the proposed multicritical DQC point. Here we introduce the SU($N$) $X$-$Q$ model, a sign-problem-free Hamiltonian that, like the $J$-$Q$ model, exhibits a direct AFM--VBS transition but differs in the operator that stabilizes the AFM state; the $X$ term consists of permutation operators on second-neighbor sites instead of the standard exchange $J$. The $X$-$Q$ model supports an AFM--VBS transition at arbitrary $N$, whereas the $N>4$ $J$-$Q$ models only do if other interactions are added \cite{Beach_PRB_2009,Kaul_PRL_2012}. Contrary to expectations that increasing $N$ favors critical behavior, we find the opposite trend: while the transition is only weakly first-order for SU($2$), it becomes increasingly discontinuous for larger  $N$.

We attribute this unexpected behavior to the absence of emergent U(1) symmetry in the VBS order parameter for large $N$ (in practice
all $N > 2$). Unlike the singlet projectors of the $J$-$Q$ model, the $X$ permutations do not induce resonances between local VBS orientations, thus not enabling global rotations of the order parameter. This insight allows us to explain why the large-$N$ $X$-$Q$ model does not flow to the same non-compact CP$^{N-1}$ continuum description \cite{Senthil_Science_2004,Senthil_PRB_2004,Dyer_JHEP_2015} as the extended $J$-$Q$ and $J_1$-$J_2$ Heisenberg models
\cite{Kaul_PRL_2012,Block_PRL_2013}, instead freezing into a trivial columnar VBS in a first-order transition.

\emph{Model and Methods.}---We define the SU($N$) square-lattice $X$-$Q$ model in the standard way \cite{Read_PRL_1989,Read_PRB_1990} in which the two sublattices carry the fundamental and conjugate SU($N$) representations, corresponding to Young tableaux with one box and a column of $N-1$ boxes respectively. This structure allows nearest neighbors to form SU($N$) singlets of $N$ ``colors''; see End Matter for SU($N$) algebra details. The Hamiltonian is
\begin{equation}\label{Eq:Ham}
    H = -\frac{X}{N^2}\sum_{\bracket{\bracket{ijkl}}}\Pi_{il}\Pi_{jk} - Q\sum_{\bracket{ijkl}}P_{ij}P_{kl},
\end{equation}
where $\langle ijkl \rangle$ indicates that $P_{ij}$ (singlet projectors) act on nearest-neighbor bonds of $2\times2$ plaquettes and $\langle\langle ijkl \rangle\rangle$ that $\Pi_{ij}$ (permutation operators) act on same-sublattice second-neighbor bonds on a periodic $L\times L$ lattice. The $1/N^2$ factor in front of the X term is a convention choice, motivated by $\frac{1}{N}\Pi_{ij}$ being the conventional ferromagnetic exchange. The operators can be written in terms of the SU($N$) generators $T^a$ as
\begin{subequations}\label{Eq:Ops}
\begin{eqnarray}
    P_{ij} = \frac{1}{N^2} - \frac{2}{N}T^a_i\bar{T}_j^a, \label{projdef} \\
    \frac{1}{N}\Pi_{ij} = \frac{1}{N^2} + \frac{2}{N}T^a_iT^a_j, \label{permdef}
\end{eqnarray}
\end{subequations}
Where $\bar{T}^a = -(T^a)^*$ denotes the conjugate representation generators. Two-site SU($N$) singlets are then superpositions
of all same-color configurations;
\begin{equation}
\ket{s} = \frac{1}{\sqrt{N}}\sum_{\alpha}\ket{\alpha_i\bar{\alpha}_j},~~ \alpha \in \{1,\ldots,N\}.
\end{equation}
where $\bar{\alpha}_j$ indicates the state on site $j$ transforms under the conjugate representation. To be precise, the action of each operator on color basis states $\alpha \in 1\ldots N$ is
\begin{subequations}\label{Eq:Actions}
\begin{eqnarray}
    P_{ij}\ket{\alpha_i\bar{\beta}_j} = \frac{1}{\sqrt{N}}\delta_{\alpha\beta}\ket{s} \\
    \frac{1}{N}\Pi_{ij}\ket{\alpha_i\beta_j} = \frac{1}{N}\ket{\beta_i\alpha_j}.
\end{eqnarray}
\end{subequations}

The $X$ term permutes colors on the same sublattice, favoring color alignment within each sublattice and stabilizing generalized AFM order. The $Q$ term favors singlet formation on nearest-neighbor bonds and couples the two sublattices, promoting VBS order. The model is sign-problem free for all $N$ and we study it using the stochastic series expansion (SSE) QMC method with operator-loop updates \cite{Sandvik_PRB_1999,Sandvik_AIP_2010} at temperatures sufficiently low ($T \propto L^{-1}$) to converge to the ground state. For efficient sampling near the strongly first-order $N>2$ transitions, we employ quantum parallel tempering \cite{Sengupta_PRB_2002,Hukushima_JPhysSoc_1996} in the coupling ratio $X/Q$. 

\begin{figure*}[t] 
\centering
\includegraphics[width=\textwidth]{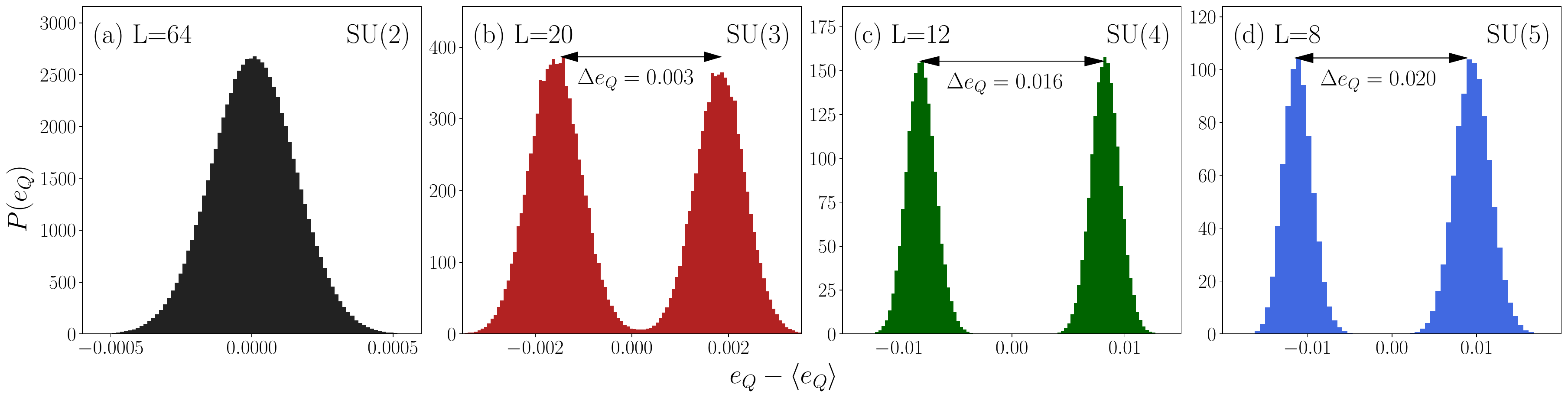}
\caption{$Q$-energy histograms relative to the mean for the SU($N$) models. For $N=2$ in (a), the distribution for $L=64$ has a single
peak, while for $N = 3,4,5$ in (b), (c), and (d), respectively, the large peak splitting for the smaller sizes (which are the largest
systems studied) signals two-phase coexistence. The indicated peak splitting is almost converged versus $L$.}
\label{fig:Energy}
\end{figure*}

We detect AFM order via the staggered magnetization $M_s$, while VBS order is probed through its columnar dimer (singlet) pattern. We define these order parameters for SU($N$) spins in terms of the $N-1$ Cartan matrices $H^a$ (listed in the Appendix)
\begin{subequations}\label{Eq:Order}
\begin{eqnarray}
    m_s^a &=& \frac{1}{L^2}\sum_{i,\alpha} H^a_{\alpha\alpha}n_{i\alpha}, \label{msdef} \\
    D_\mu^a &=& \frac{1}{L^2}\sum_{i,\alpha} (-1)^\mu H^a_{\alpha\alpha}n_{i,\alpha}H^a_{\alpha\alpha}n_{i+\hat{\mu},\alpha}, \label{ddef}
\end{eqnarray}
\end{subequations}
where $\alpha = 1 \ldots N$ is for the $N$ colors, $n_{i,\alpha} = 1$ if color $\alpha$ is on site $i$, and $\mu = x,y$ denotes
$x$ and $y$ orientations. The simulations do not break any symmetries, and we study the squared magnitudes of the order parameters,
defining $M_s^2 = \langle \sum_a (m_s^a)^2\rangle$ and $|\psi|^2=\langle D_x^2+D_y^2\rangle$, with $D_\mu = \sum_a D^a_\mu$.
To test for the emergence of U(1) symmetry at the transition, we evaluate the anisotropy $\phi_4 = \bracket{\cos(4\theta)}$,
where $\theta$ is the angle corresponding to $\psi = D_x+iD_y$.

For a standard diagnostic of the phase transitions, we define the Binder cumulant for $\mathcal{O} = M_s$ or $\mathcal{O} = \psi$ as
\begin{equation}\label{Eq:Binder}
    U_2 = 1 - \frac{N-1}{N+1}\frac{\bracket{\mathcal{O}^4}}{\bracket{\mathcal{O}^2}^2},
\end{equation}
where the $N$-dependent prefactor accounts for the multi-component nature of the SU($N$) order parameter and reduces to the standard value of $1/3$ for SU(2). This factor ensures $U_2 \to 0$ in the disordered phase where $\bracket{\mathcal{O}^2} \to 0$ when $L\to \infty$. To more directly probe first-order behavior, we will also use the probability distribution $P(e_Q)$ of the $Q$ part of the energy

\emph{Results.}---We first compare the SU(2) and SU(3) models. Fig.~\ref{fig:Binder}(a) shows positive-definite cumulants for SU(2) systems up to size $L=256$, and fitted curves produce a crossing point at $(X/Q)_c =  0.1152(2)$. In sharp contrast, for SU(3) systems the cumulants in Fig.~\ref{fig:Binder}(b) show negative dips that deepen in proportion to the space-time volume $L^3$, which is a hallmark of a first-order transition \cite{Vollmayr_ZPB_1993,Iino_JPhysSoc_2019}. Tunneling between the different phases  near the transition quickly becomes intractable, and we have only equilibrated systems up to size $L=20$. The drifts of the  
crossings and peaks indicate $(X/Q)_c=5.223(2)$. In Fig.~\ref{fig:Binder}(c), $L \to \infty$  extrapolations of the order parameters show non-vanishing values at the transition, thus demonstrating phase coexistence. For SU(2), the absence of negative cumulant and the very small extrapolated order parameters puts the transition in the category of very weak first-order transitions. The extrapolated coexistence is robust to fitting windows and functions, supporting the same weakly first order behavior as the $J$-$Q$ model \cite{Takahashi_2024}. The very different behaviors for SU(3) suggest that this model is far from any critical point.

\begin{figure*}[t] 
  \centering
  \includegraphics[width=\textwidth]{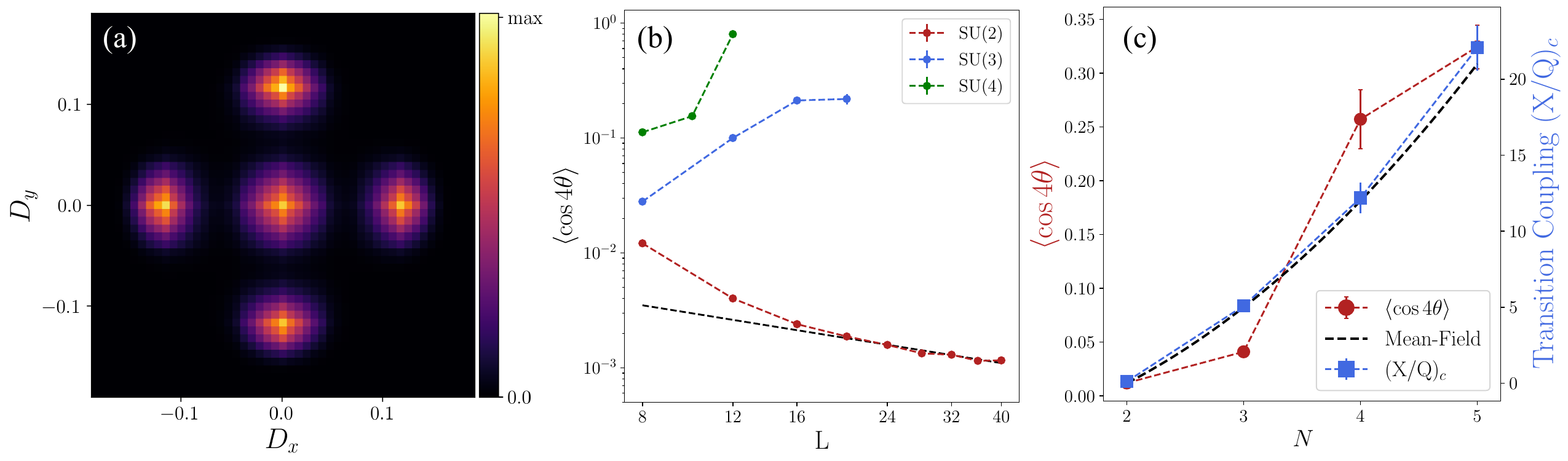}
  \caption{(a) Distribution of the VBS order parameter $\psi=D_x+iD_y$ at the transition point for the SU(3) $L=20$ system. Peaks at $(D_x,D_y) = (\pm D,0),(0,\pm D)$ correspond to the VBS order and the peak at $|D|=0$ arises from coexistence with the AFM phase. (b) The  anisotropy parameter increases with $L$ for SU(3) and SU(4) but decreases for SU($2$), with the line indicating the expected $L^{-\mu}$ form with $\mu = 0.72$ \cite{Takahashi_2024} for a system sufficiently close to the DQC point. (c) Anisotropy parameter (red, left axis) for $L=8$ at the estimated $L \to \infty$ transition points (blue, right axis) for SU(2)-SU(5). Transition points agree well with a mean-field calculation (black dashed curve).}
  \label{fig:Anisotropy}
\end{figure*}

Fig.~\ref{fig:Energy} shows the $Q$ energy-per-site distribution $P(e_Q)$ at the transition for $N=2$-$5$ computed for sizes $L=64,20,12,8$, respectively. At a quantum phase transition ($T=0$), the total energy $e = \bracket{H}/L^2$ is analogous to the free energy at a classical $T>0$ transition and $P(e)$ is therefore always singly-peaked. The distribution $P(e_Q)$ [or, equivalently, $P(e_X)$] can be used to detect a discontinuity analogously to the internal energy classically. For $N>2$, the distributions in Fig.~\ref{fig:Energy} are bimodal; a defining signature of phase coexistence at a first-order transition. There is no evidence of bimodality for SU(2), though we do expect the distribution to eventually split for much larger $L$, given the weak first-order transition demonstrated in Fig.~\ref{fig:Binder}. We find that the peak separation $\Delta e_Q$ increases with $N$, demonstrating that the transition becomes more strongly first-order, in contrast to the behavior of the SU($N$) $J$-$Q$ models
\cite{Lou_PRB_2009,Kaul_PRL_2012}.

The strongly first-order transitions in the SU($N>2$) models are associated with the absence of emergent U($1$) symmetry; the lattice anisotropy remains relevant. The DQC scenario requires that this anisotropy be ``dangerously irrelevant'', i.e., the presumed continuum critical point is stable in the presence of the lattice only if the U($1$) symmetry is emergent for $L \to \infty$. The relevance of the lattice in the VBS phase is associated with a length scale different from the conventional correlation length, and this length scale is also related to the scale of spinon deconfinement in the neighborhood of the DQC transition \cite{Senthil_PRB_2004,Shao_Science_2016}. The emergent U($1$) symmetry is well known in the case of extended $J$-$Q$ models for any $N$ \cite{Sandvik_PRL_2007,Lou_PRB_2009,Jiang_JStatMech_2008} and, for the special case of $N=2$, was later shown to be further enhanced to SO(5), in related 3D classical models \cite{Nahum_PRL_2015,Sreejith_PRL_2019} and in the $J$-$Q$ models \cite{Takahashi_PRR_2020,Takahashi_2024}.

The distribution of the VBS order parameter in Fig.~\ref{fig:Anisotropy}(a) reflects phase coexistence in the SU($3$) case, with only $\mathbb{Z}_4$ symmetry (spontaneously broken for $L \to \infty$) of the VBS order. An emergent U($1$) symmetry would appear as an increasingly uniform ring $|\psi| = D$ in the distribution as $L$ increases \cite{Sandvik_PRL_2007}. Fig.~\ref{fig:Anisotropy}(b) shows the size-dependent anisotropy parameter $\phi_4 = \langle \cos 4\theta\rangle$ for systems at their transition points, with $\theta$ extracted from equal-time values of $D_x,D_y$ defined in Eq.~(\ref{ddef}). For SU(2) and SU(3), the transition point is determined via the cumulant crossings and minimums respectively in Fig.~\ref{fig:Binder}(a), while for SU(4) and SU(5) we use the size-dependent point where the energy distribution [Fig.~\ref{fig:Energy}] is maximally bimodal. If anisotropy remains, $\phi_4$ will converge to a nonzero value, as it does for $N > 2$, while  $\phi_4 \to 0$ if there is emergent U($1$) symmetry. In the case of SU($2$), we do find emergent U($1$) symmetry for the system sizes studied here [presumably also SO(5) but we have not investigated it here]. The size dependence $\phi_4 \sim L^{-\mu}$ reflects the scaling dimension of the perturbation, which was determined by much more comprehensive results for the $J$-$Q$ model \cite{Takahashi_2024}; $\mu \approx 0.72$. The $L \agt 20$ $X$-$Q$ results in Fig.~\ref{fig:Anisotropy}(b) are consistent with the same $\mu$.

\begin{figure}[t] 
\includegraphics[width=70mm]{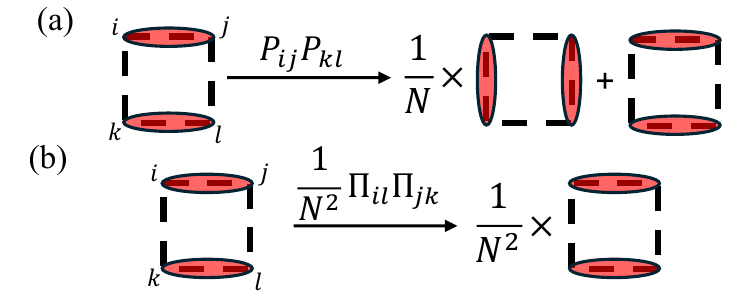}
\caption{Action and matrix elements of the (a) $Q$ singlet-projectors and (b) $X$ permutations on a single VBS plaquette. The $X$ term does not mix VBS orientations. The $Q$ term rotates VBS orientations, leading to U(1) resonances.}
\label{fig:Cartoon}
\end{figure}

In past studies of strongly first-order transitions in SU($2$) $J$-$Q$ type models, the singlet projectors were arranged to produce a staggered VBS on square \cite{Sen_PRB_2010} and honeycomb \cite{Banerjee_PRB_2011} lattices. Such VBSs do not support local singlet rotations; thus U(1) symmetry cannot emerge and these models host conventional strongly first-order transitions not influenced by any nearby critical point. In the $X$-$Q$ model, the emergent U(1) symmetry is violated in a more subtle manner, which we can relate to the way an $X$ term acts on a local $2\times 2$ plaquette with two parallel singlets, illustrated and compared with a $Q$ term in Fig.~\ref{fig:Cartoon}. While the $Q$ term (and the not shown $J$ terms) flips the orientation of the singlets, the $X$ term does not. Thus, a series of $Q$ interactions can gradually rotate the global VBS order, but the $X$ terms do not induce these fluctuations that are necessary for U($1$) symmetry to emerge at a DQC point.

\emph{Large-N proof}.---We prove a first-order transition in the $X$-$Q$ model for large $N$ as follows: in the overcomplete bipartite SU($N$) valence-bond (singlet) basis, the diagonal and off-diagonal matrix elements of the projectors $P_{ij}$ are $1$ and $1/N$, respectively. The same-sublattice permutation operators $\Pi/N$ only have off-diagonal matrix elements $1/N$. In a perfect columnar VBS, the $Q$ term acting on a plaquette in Fig.~\ref{fig:Cartoon}(a) flips the singlets with matrix element $1/N$ (from an off-diagonal followed by a diagonal operation), while the $X$ term in Fig.~\ref{fig:Cartoon}(b) leaves the state unchanged with matrix element $1/N^2$. For $N\to\infty$, only the diagonal part survives and the columnar VBS pattern is the ground state of the $Q$ interaction. To destroy the order, the $X$ interaction has to come with a factor of order $N^2$ to compete with the diagonal $Q$ interaction, i.e., $(X/Q)_c\sim N^2$. A quantitative mean field theory (presented in End Matter) based on these large-$N$ arguments gives $(X/Q)_c = N^2 -4$, which agrees very well with the computed data points in Fig.~\ref{fig:Anisotropy}(c) even for small $N$.

Since the off-diagonal $Q$ terms are suppressed by $1/N$ and the $X$ interaction cannot rotate singlets, there can be no emergent U(1) symmetry and the fundamental prerequisite from DQC is missing. Fig.~\ref{fig:Anisotropy}(c) also shows $\phi_4$ for fixed $L=8$, demonstrating rapid loss of U(1) fluctuations with increasing $N$. In contrast to the $X$-$Q$ model, for the extended Heisenberg model with singlet projectors of strength $J_1$ on nearest neighbors and permutations of strength $J_2$ on next-nearest neighbors, the ability of the $J_2$ permutations to rotate singlets enables the emergent U(1) symmetry and critical behavior in quantitative agreement with a $1/N$ expansion for the $CP^{N-1}$ field theory was found \cite{Kaul_PRL_2012,Dyer_JHEP_2015}.

\emph{Discussion}.---The first order mechanism is at play in the $X$-$Q$ model already for $N=3$, where $(X/Q)_c = 5.223(2)$ and the anisotropy $\phi_4$ increases with $L$ starting from the smallest systems in Fig.~\ref{fig:Anisotropy}(b). In contrast, in the case of $N=2$, $\phi_4$ decreases, following closely the expected \cite{Takahashi_2024} power-law at a DQC point. This system, with its rather small $(X/Q)_c = 0.1152(2)$ value [i.e. the U(1) enabling $Q$ term dominates] is near-DQC and the picture of a dangerously irrelevant lattice perturbation applies. The emergent SO($5$) symmetry for (only) $N=2$ \cite{Senthil_PRB_2006} is a possible additional contributing factor for this case being special. Given that the $Q$-only model is already quite close to critical \cite{Sandvik_PRL_2007}, it also has significant SO($5$) character and the weak $X$ perturbation retains that additional channel of fluctuations that is absent for $N>2$.

In both the $J$-$Q$ and SU($2$) $X$-$Q$ models, in the scenario of Ref. \cite{Takahashi_2024} the primary reason for the system not being exactly at the DQC point is not a lack of emergent symmetry (which is a secondary effect) but the presence of an SO($5$) singlet perturbation that is not tuned away by only changing a ratio such as $J/Q$. The multicritical point, which is supported also by CFT work \cite{Chester_PRL_2024}, requires additional interactions. However, reaching the singular point without a QMC sign problem may not be possible, or at the very least requires interactions not yet considered. It will certainly be interesting to study a combined $J_1$-$J_2$-$X$-$Q$ model, similar to the $J$-$Q$ model with higher-order $Q$ terms in Ref. \cite{Takahashi_2024}, to potentially tune closer to the putative multicritical point. There the system was tuned from weak to moderately strong first-order transitions, with scaling behavior on the coexistence line consistent with the expectations with the exponents extracted in other ways. The effects of $X$ interactions for both small and large $N$ should give further insights into the nature of perturbed DQC transitions. 

\begin{acknowledgments}
\emph{Acknowledgements.}---We would like to thank Kedar Damle for discussions. This research was supported
by the Simons Foundation under Grant No. 511064 (R.F and A.W.S) and by a Fellowship from the John Simon Guggenheim Memorial Foundation
(A.W.S.). The numerical calculations were carried out on the Shared
Computing Cluster managed by Boston University’s Research Computing Services.
\end{acknowledgments}

\bibliography{references}

@article{DEmidio_PRL_2024,
  title = {{Entanglement Entropy and Deconfined Criticality: Emergent SO(5) Symmetry and Proper Lattice Bipartition}},
  author = {D'Emidio, Jonathan and Sandvik, Anders W.},
  journal = {Phys. Rev. Lett.},
  volume = {133},
  issue = {16},
  pages = {166702},
  numpages = {7},
  year = {2024},
  month = {Oct},
  publisher = {American Physical Society},
  doi = {10.1103/PhysRevLett.133.166702},
  url = {https://link.aps.org/doi/10.1103/PhysRevLett.133.166702}
}

@article{Zhao_PRL_2022,
  title = {Scaling of Entanglement Entropy at Deconfined Quantum Criticality},
  author = {Zhao, Jiarui and Wang, Yan-Cheng and Yan, Zheng and Cheng, Meng and Meng, Zi Yang},
  journal = {Phys. Rev. Lett.},
  volume = {128},
  issue = {1},
  pages = {010601},
  numpages = {6},
  year = {2022},
  month = {Jan},
  publisher = {American Physical Society},
  doi = {10.1103/PhysRevLett.128.010601},
  url = {https://link.aps.org/doi/10.1103/PhysRevLett.128.010601}
}

@article{Chester_PRL_2024,
  title = {Bootstrapping Deconfined Quantum Tricriticality},
  author = {Chester, Shai M. and Su, Ning},
  journal = {Phys. Rev. Lett.},
  volume = {132},
  issue = {11},
  pages = {111601},
  numpages = {7},
  year = {2024},
  month = {Mar},
  publisher = {American Physical Society},
  doi = {10.1103/PhysRevLett.132.111601},
  url = {https://link.aps.org/doi/10.1103/PhysRevLett.132.111601}
}

@article{Kaul_PRB_2007,
  title = {Hole dynamics in an antiferromagnet across a deconfined quantum critical point},
  author = {Kaul, Ribhu K. and Kolezhuk, Alexei and Levin, Michael and Sachdev, Subir and Senthil, T.},
  journal = {Phys. Rev. B},
  volume = {75},
  issue = {23},
  pages = {235122},
  numpages = {17},
  year = {2007},
  month = {Jun},
  publisher = {American Physical Society},
  doi = {10.1103/PhysRevB.75.235122},
  url = {https://link.aps.org/doi/10.1103/PhysRevB.75.235122}
}

@article{Anderson_Science_1987,
  author  = {P. W. Anderson},
  title   = {{The Resonating Valence Bond State in {La$_2$CuO$_4$} and Superconductivity}},
  journal = {Science},
  volume  = {235},
  number  = {4793},
  pages   = {1196-1198},
  year    = {1987},
  doi     = {10.1126/science.235.4793.1196},
  url     = {https://www.science.org/doi/abs/10.1126/science.235.4793.1196}
}

@article{Chakravarty_PRB_1989,
  title = {{Two-dimensional quantum {Heisenberg} antiferromagnet at low temperatures}},
  author = {Chakravarty, Sudip and Halperin, Bertrand I. and Nelson, David R.},
  journal = {Phys. Rev. B},
  volume = {39},
  issue = {4},
  pages = {2344--2371},
  numpages = {0},
  year = {1989},
  month = {Feb},
  publisher = {American Physical Society},
  doi = {10.1103/PhysRevB.39.2344},
  url = {https://link.aps.org/doi/10.1103/PhysRevB.39.2344}
}

@article{Dagotto_PRL_1989,
  title = {{Phase diagram of the frustrated spin-1/2 {Heisenberg} antiferromagnet in 2 dimensions}},
  author = {Dagotto, Elbio and Moreo, Adriana},
  journal = {Phys. Rev. Lett.},
  volume = {63},
  issue = {19},
  pages = {2148--2151},
  numpages = {0},
  year = {1989},
  month = {Nov},
  publisher = {American Physical Society},
  doi = {10.1103/PhysRevLett.63.2148},
    url ={https://link.aps.org/doi/10.1103/PhysRevLett.63.2148}
}

@article{Inui_PRB_1988,
  title = {{Coexistence of antiferromagnetism and superconductivity in a mean-field theory of high-${T}_{c}$ superconductors}},
  author = {Inui, Masahiko and Doniach, Sebastian and Hirschfeld, Peter J. and Ruckenstein, Andrei E.},
  journal = {Phys. Rev. B},
  volume = {37},
  issue = {4},
  pages = {2320--2323},
  numpages = {0},
  year = {1988},
  month = {Feb},
  publisher = {American Physical Society},
  doi = {10.1103/PhysRevB.37.2320},
  url = {https://link.aps.org/doi/10.1103/PhysRevB.37.2320}
}

@article{Manousakis_RevModPhys_1991,
  title = {{The spin-\textonehalf{} {Heisenberg} antiferromagnet on a square lattice and its application to the cuprous oxides}},
  author = {Manousakis, Efstratios},
  journal = {Rev. Mod. Phys.},
  volume = {63},
  issue = {1},
  pages = {1--62},
  numpages = {0},
  year = {1991},
  month = {Jan},
  publisher = {American Physical Society},
  doi = {10.1103/RevModPhys.63.1},
  url = {https://link.aps.org/doi/10.1103/RevModPhys.63.1}
}

@article{Chubukov_PRB_1994,
  title = {{Theory of two-dimensional quantum {Heisenberg} antiferromagnets with a nearly critical ground state}},
  author = {Chubukov, Andrey V. and Sachdev, Subir and Ye, Jinwu},
  journal = {Phys. Rev. B},
  volume = {49},
  issue = {17},
  pages = {11919--11961},
  numpages = {0},
  year = {1994},
  month = {May},
  publisher = {American Physical Society},
  doi = {10.1103/PhysRevB.49.11919},
  url = {https://link.aps.org/doi/10.1103/PhysRevB.49.11919}
}

@incollection{Sachdev_WS_1995,
  author    = {Subir Sachdev},
  title     = {{Low-dimensional quantum field theories of quantum antiferromagnets}},
  booktitle = {Low Dimensional Quantum Field Theories for Condensed Matter Physicists},
  editor    = {Y. Lu and S. Lundqvist and G. Morandi},
  publisher = {World Scientific},
  address   = {Singapore},
  year      = {1995},
  pages     = {381--432}
}

@article{Anderson_MatResBul_1973,
  title={{Resonating valence bonds: A new kind of insulator?}},
  author={Philip W. Anderson},
  journal={Materials Research Bulletin},
  year={1973},
  volume={8},
  pages={153-160},
  url={https://api.semanticscholar.org/CorpusID:135766371}
}

@article{Read_PRL_1989,
  title = {{Valence-bond and spin-{Peierls} ground states of low-dimensional quantum antiferromagnets}},
  author = {Read, N. and Sachdev, Subir},
  journal = {Phys. Rev. Lett.},
  volume = {62},
  issue = {14},
  pages = {1694--1697},
  numpages = {0},
  year = {1989},
  month = {Apr},
  publisher = {American Physical Society},
  doi = {10.1103/PhysRevLett.62.1694},
  url = {https://link.aps.org/doi/10.1103/PhysRevLett.62.1694}
}

@article{Read_PRB_1990,
  title = {{Spin-{Peierls}, valence-bond solid, and N\'eel ground states of low-dimensional quantum antiferromagnets}},
  author = {Read, N. and Sachdev, Subir},
  journal = {Phys. Rev. B},
  volume = {42},
  issue = {7},
  pages = {4568--4589},
  numpages = {0},
  year = {1990},
  month = {Sep},
  publisher = {American Physical Society},
  doi = {10.1103/PhysRevB.42.4568},
  url = {https://link.aps.org/doi/10.1103/PhysRevB.42.4568}
}

@article{Bednorz_ZPB_1986,
  author  = {J. G. Bednorz and K. A. M{\"u}ller},
  title   = {{Possible high {$T_c$} superconductivity in the {Ba--La--Cu--O} system}},
  journal = {Z. Phys. B},
  volume  = {64},
  pages   = {189--193},
  year    = {1986},
  doi     = {10.1007/BF01303701},
  url = {https://link.springer.com/article/10.1007/BF01303701#citeas}
}

@book{Auerbach1994,
  author = {Auerbach, A.},
  title = {{Interacting Electrons and Quantum Magnetism}},
  publisher = {Springer-Verlag},
  address = {New York},
  year = {1994},
  series = {Graduate Texts in Contemporary Physics},
  isbn = {978-0-387-94286-5}
}

@article{Lee_RevModPhys_2006,
  title = {{Doping a {Mott} insulator: Physics of high-temperature superconductivity}},
  author = {Lee, Patrick A. and Nagaosa, Naoto and Wen, Xiao-Gang},
  journal = {Rev. Mod. Phys.},
  volume = {78},
  issue = {1},
  pages = {17--85},
  numpages = {0},
  year = {2006},
  month = {Jan},
  publisher = {American Physical Society},
  doi = {10.1103/RevModPhys.78.17},
  url = {https://link.aps.org/doi/10.1103/RevModPhys.78.17}
}

@book{Sachdev2011,
  author    = {Sachdev, Subir},
  title     = {{Quantum Phase Transitions}},
  edition   = {2},
  publisher = {Cambridge University Press},
  address   = {Cambridge},
  year      = {2011},
  isbn      = {9780521514682},
  doi       = {10.1017/CBO9780511973765}
}

@article{Scalapino_RevModPhys_2012,
  title = {{A common thread: The pairing interaction for unconventional superconductors}},
  author = {Scalapino, D. J.},
  journal = {Rev. Mod. Phys.},
  volume = {84},
  issue = {4},
  pages = {1383--1417},
  numpages = {0},
  year = {2012},
  month = {Oct},
  publisher = {American Physical Society},
  doi = {10.1103/RevModPhys.84.1383},
  url = {https://link.aps.org/doi/10.1103/RevModPhys.84.1383}
}

@article{Sachdev_NatPhys_2008,
  author  = {Subir Sachdev},
  title   = {{Quantum magnetism and criticality}},
  journal = {Nature Physics},
  volume  = {4},
  number  = {3},
  pages   = {173--185},
  year    = {2008},
  doi     = {10.1038/nphys894}
}

@article{Haldane_PRL_1988,
  title = {{O(3) Nonlinear $\ensuremath{\sigma}$ Model and the Topological Distinction between Integer- and Half-Integer-Spin Antiferromagnets in Two Dimensions}},
  author = {Haldane, F. D. M.},
  journal = {Phys. Rev. Lett.},
  volume = {61},
  issue = {8},
  pages = {1029--1032},
  numpages = {0},
  year = {1988},
  month = {Aug},
  publisher = {American Physical Society},
  doi = {10.1103/PhysRevLett.61.1029},
  url = {https://link.aps.org/doi/10.1103/PhysRevLett.61.1029}
}

@article{Senthil_Science_2004,
  author  = {Senthil, T. and Vishwanath, A. and Balents, L. and Sachdev, S. and Fisher, M. P. A.},
  title   = {{Deconfined Quantum Critical Points}},
  journal = {Science},
  volume  = {303},
  number  = {5663},
  pages   = {1490--1494},
  year    = {2004},
  doi     = {10.1126/science.1091806}
}

@article{Levin_PRB_2004,
  title = {{Deconfined quantum criticality and N\'eel order via dimer disorder}},
  author = {Levin, Michael and Senthil, T.},
  journal = {Phys. Rev. B},
  volume = {70},
  issue = {22},
  pages = {220403},
  numpages = {4},
  year = {2004},
  month = {Dec},
  publisher = {American Physical Society},
  doi = {10.1103/PhysRevB.70.220403},
  url = {https://link.aps.org/doi/10.1103/PhysRevB.70.220403}
}

@article{Senthil_PRB_2004,
  title = {{Quantum criticality beyond the Landau-Ginzburg-Wilson paradigm}},
  author = {Senthil, T. and Balents, Leon and Sachdev, Subir and Vishwanath, Ashvin and Fisher, Matthew P. A.},
  journal = {Phys. Rev. B},
  volume = {70},
  issue = {14},
  pages = {144407},
  numpages = {33},
  year = {2004},
  month = {Oct},
  publisher = {American Physical Society},
  doi = {10.1103/PhysRevB.70.144407},
  url = {https://link.aps.org/doi/10.1103/PhysRevB.70.144407}
}

@article{Senthil_PRB_2006,
  title = {{Competing orders, nonlinear sigma models, and topological terms in quantum magnets}},
  author = {Senthil, T. and Fisher, Matthew P. A.},
  journal = {Phys. Rev. B},
  volume = {74},
  issue = {6},
  pages = {064405},
  numpages = {11},
  year = {2006},
  month = {Aug},
  publisher = {American Physical Society},
  doi = {10.1103/PhysRevB.74.064405},
  url = {https://link.aps.org/doi/10.1103/PhysRevB.74.064405}
}

@article{Sandvik_PRL_2007,
  title = {{Evidence for Deconfined Quantum Criticality in a Two-Dimensional {Heisenberg} Model with Four-Spin Interactions}},
  author = {Sandvik, Anders W.},
  journal = {Phys. Rev. Lett.},
  volume = {98},
  issue = {22},
  pages = {227202},
  numpages = {4},
  year = {2007},
  month = {Jun},
  publisher = {American Physical Society},
  doi = {10.1103/PhysRevLett.98.227202},
  url = {https://link.aps.org/doi/10.1103/PhysRevLett.98.227202}
}

@article{Nahum_PRX_2015,
  title = {{Deconfined Quantum Criticality, Scaling Violations, and Classical Loop Models}},
  author = {Nahum, Adam and Chalker, J. T. and Serna, P. and Ortu\~no, M. and Somoza, A. M.},
  journal = {Phys. Rev. X},
  volume = {5},
  issue = {4},
  pages = {041048},
  numpages = {21},
  year = {2015},
  month = {Dec},
  publisher = {American Physical Society},
  doi = {10.1103/PhysRevX.5.041048},
  url = {https://link.aps.org/doi/10.1103/PhysRevX.5.041048}
}

@article{Sreejith_PRL_2019,
  title = {{Emergent SO(5) Symmetry at the Columnar Ordering Transition in the Classical Cubic Dimer Model}},
  author = {Sreejith, G. J. and Powell, Stephen and Nahum, Adam},
  journal = {Phys. Rev. Lett.},
  volume = {122},
  issue = {8},
  pages = {080601},
  numpages = {6},
  year = {2019},
  month = {Feb},
  publisher = {American Physical Society},
  doi = {10.1103/PhysRevLett.122.080601},
  url = {https://link.aps.org/doi/10.1103/PhysRevLett.122.080601}
}

@article{Vollmayr_ZPB_1993,
  author  = {K. Vollmayr and J. D. Reger and M. Scheucher and K. Binder},
  title   = {{Finite size effects at thermally-driven first order phase transitions: A phenomenological theory of the order parameter distribution}},
  journal = {Z. Phys. B},
  volume  = {91},
  pages   = {113--125},
  year    = {1993},
  doi     = {10.1007/BF01316713}
}

@article{Iino_JPhysSoc_2019,
author = {Iino ,Shumpei and Morita ,Satoshi and Kawashima ,Naoki and Sandvik ,Anders W.},
title = {{Detecting Signals of Weakly First-order Phase Transitions in Two-dimensional Potts Models}},
journal = {J. Phys. Soc. Jpn.},
volume = {88},
number = {3},
pages = {034006},
year = {2019},
doi = {10.7566/JPSJ.88.034006},
URL = {https://doi.org/10.7566/JPSJ.88.034006}
}

@article{Takahashi_PRR_2020,
  title = {{Valence-bond solids, vestigial order, and emergent SO(5) symmetry in a two-dimensional quantum magnet}},
  author = {Takahashi, Jun and Sandvik, Anders W.},
  journal = {Phys. Rev. Res.},
  volume = {2},
  issue = {3},
  pages = {033459},
  numpages = {29},
  year = {2020},
  month = {Sep},
  publisher = {American Physical Society},
  doi = {10.1103/PhysRevResearch.2.033459},
  url = {https://link.aps.org/doi/10.1103/PhysRevResearch.2.033459}
}

@article{Jiang_JStatMech_2008,
doi = {10.1088/1742-5468/2008/02/P02009},
url = {https://doi.org/10.1088/1742-5468/2008/02/P02009},
year = {2008},
month = {feb},
publisher = {},
volume = {2008},
number = {02},
pages = {P02009},
author = {Jiang, F-J and Nyfeler, M and Chandrasekharan, S and Wiese, U-J},
title = {{From an antiferromagnet to a valence bond solid: evidence for a first-order phase
transition}},
journal = {J. Stat. Mech.: Theory Exp.}
}

@article{Lou_PRB_2009,
  title = {{Antiferromagnetic to valence-bond-solid transitions in two-dimensional {$\text{SU}(N)$ Heisenberg} models with multispin interactions}},
  author = {Lou, Jie and Sandvik, Anders W. and Kawashima, Naoki},
  journal = {Phys. Rev. B},
  volume = {80},
  issue = {18},
  pages = {180414},
  numpages = {4},
  year = {2009},
  month = {Nov},
  publisher = {American Physical Society},
  doi = {10.1103/PhysRevB.80.180414},
  url = {https://link.aps.org/doi/10.1103/PhysRevB.80.180414}
}

@article{Nahum_PRL_2015,
  title = {{Emergent SO(5) Symmetry at the N\'eel to Valence-Bond-Solid Transition}},
  author = {Nahum, Adam and Serna, P. and Chalker, J. T. and Ortu\~no, M. and Somoza, A. M.},
  journal = {Phys. Rev. Lett.},
  volume = {115},
  issue = {26},
  pages = {267203},
  numpages = {5},
  year = {2015},
  month = {Dec},
  publisher = {American Physical Society},
  doi = {10.1103/PhysRevLett.115.267203},
  url = {https://link.aps.org/doi/10.1103/PhysRevLett.115.267203}
}

@article{Melko_PRL_2008,
  title = {{Scaling in the Fan of an Unconventional Quantum Critical Point}},
  author = {Melko, Roger G. and Kaul, Ribhu K.},
  journal = {Phys. Rev. Lett.},
  volume = {100},
  issue = {1},
  pages = {017203},
  numpages = {4},
  year = {2008},
  month = {Jan},
  publisher = {American Physical Society},
  doi = {10.1103/PhysRevLett.100.017203},
  url = {https://link.aps.org/doi/10.1103/PhysRevLett.100.017203}
}

@article{Kaul_PRB_2011,
  title = {{Quantum criticality in SU(3) and SU(4) antiferromagnets}},
  author = {Kaul, Ribhu K.},
  journal = {Phys. Rev. B},
  volume = {84},
  issue = {5},
  pages = {054407},
  numpages = {8},
  year = {2011},
  month = {Aug},
  publisher = {American Physical Society},
  doi = {10.1103/PhysRevB.84.054407},
  url = {https://link.aps.org/doi/10.1103/PhysRevB.84.054407}
}

@article{Kaul_PRL_2012,
  title = {{Lattice Model for the $\mathrm{SU}(N)$ N\'eel to Valence-Bond Solid Quantum Phase Transition at Large $N$}},
  author = {Kaul, Ribhu K. and Sandvik, Anders W.},
  journal = {Phys. Rev. Lett.},
  volume = {108},
  issue = {13},
  pages = {137201},
  numpages = {5},
  year = {2012},
  month = {Mar},
  publisher = {American Physical Society},
  doi = {10.1103/PhysRevLett.108.137201},
  url = {https://link.aps.org/doi/10.1103/PhysRevLett.108.137201}
}

@article{Block_PRL_2013,
  title = {{Fate of $\mathbb{C}{\mathbb{P}}^{N\ensuremath{-}1}$ Fixed Points with $q$ Monopoles}},
  author = {Block, Matthew S. and Melko, Roger G. and Kaul, Ribhu K.},
  journal = {Phys. Rev. Lett.},
  volume = {111},
  issue = {13},
  pages = {137202},
  numpages = {5},
  year = {2013},
  month = {Sep},
  publisher = {American Physical Society},
  doi = {10.1103/PhysRevLett.111.137202},
  url = {https://link.aps.org/doi/10.1103/PhysRevLett.111.137202}
}

@article{Shao_Science_2016,
author = {Hui Shao  and Wenan Guo  and Anders W. Sandvik },
title = {{Quantum criticality with two length scales}},
journal = {Science},
volume = {352},
number = {6282},
pages = {213-216},
year = {2016},
doi = {10.1126/science.aad5007},
URL = {https://www.science.org/doi/abs/10.1126/science.aad5007}}

@article{Kuklov_PRL_2008,
  title = {{Deconfined Criticality: Generic First-Order Transition in the SU(2) Symmetry Case}},
  author = {Kuklov, A. B. and Matsumoto, M. and Prokof'ev, N. V. and Svistunov, B. V. and Troyer, M.},
  journal = {Phys. Rev. Lett.},
  volume = {101},
  issue = {5},
  pages = {050405},
  numpages = {4},
  year = {2008},
  month = {Aug},
  publisher = {American Physical Society},
  doi = {10.1103/PhysRevLett.101.050405},
  url = {https://link.aps.org/doi/10.1103/PhysRevLett.101.050405}
}

@article{Chen_PRL_2013,
  title = {{Deconfined Criticality Flow in the {Heisenberg} Model with Ring-Exchange Interactions}},
  author = {Chen, Kun and Huang, Yuan and Deng, Youjin and Kuklov, A. B. and Prokof'ev, N. V. and Svistunov, B. V.},
  journal = {Phys. Rev. Lett.},
  volume = {110},
  issue = {18},
  pages = {185701},
  numpages = {5},
  year = {2013},
  month = {May},
  publisher = {American Physical Society},
  doi = {10.1103/PhysRevLett.110.185701},
  url = {https://link.aps.org/doi/10.1103/PhysRevLett.110.185701}
}

@article{Sandvik_PRL_2010,
  title = {{Continuous Quantum Phase Transition between an Antiferromagnet and a Valence-Bond Solid in Two Dimensions: Evidence for Logarithmic Corrections to Scaling}},
  author = {Sandvik, Anders W.},
  journal = {Phys. Rev. Lett.},
  volume = {104},
  issue = {17},
  pages = {177201},
  numpages = {4},
  year = {2010},
  month = {Apr},
  publisher = {American Physical Society},
  doi = {10.1103/PhysRevLett.104.177201},
  url = {https://link.aps.org/doi/10.1103/PhysRevLett.104.177201}
}

@article{Zhao_PRL_2020,
  title = {{Multicritical Deconfined Quantum Criticality and Lifshitz Point of a Helical Valence-Bond Phase}},
  author = {Zhao, Bowen and Takahashi, Jun and Sandvik, Anders W.},
  journal = {Phys. Rev. Lett.},
  volume = {125},
  issue = {25},
  pages = {257204},
  numpages = {7},
  year = {2020},
  month = {Dec},
  publisher = {American Physical Society},
  doi = {10.1103/PhysRevLett.125.257204},
  url = {https://link.aps.org/doi/10.1103/PhysRevLett.125.257204}
}

@article{Gong_PRL_2014,
  title = {{Plaquette Ordered Phase and Quantum Phase Diagram in the {Spin-$\frac{1}{2}$ ${J}_{1}\text{\ensuremath{-}}{J}_{2}$} Square {Heisenberg} Model}},
  author = {Gong, Shou-Shu and Zhu, Wei and Sheng, D. N. and Motrunich, Olexei I. and Fisher, Matthew P. A.},
  journal = {Phys. Rev. Lett.},
  volume = {113},
  issue = {2},
  pages = {027201},
  numpages = {5},
  year = {2014},
  month = {Jul},
  publisher = {American Physical Society},
  doi = {10.1103/PhysRevLett.113.027201},
  url = {https://link.aps.org/doi/10.1103/PhysRevLett.113.027201}
}

@article{Morita_JPhysSoc_2015,
author = {Morita ,Satoshi and Kaneko ,Ryui and Imada ,Masatoshi},
title = {{Quantum Spin Liquid in {Spin 1/2 J1–J2 Heisenberg} Model on Square Lattice: Many-Variable Variational Monte Carlo Study Combined with Quantum-Number Projections}},
journal = {J. Phys. Soc. Jpn.},
volume = {84},
number = {2},
pages = {024720},
year = {2015},
doi = {10.7566/JPSJ.84.024720},
URL = { https://doi.org/10.7566/JPSJ.84.024720
}}

@article{Wang_PRL_2018,
  title = {{Critical Level Crossings and Gapless Spin Liquid in the Square-Lattice {Spin-$1/2$ ${J}_{1}\ensuremath{-}{J}_{2}$ Heisenberg} Antiferromagnet}},
  author = {Wang, Ling and Sandvik, Anders W.},
  journal = {Phys. Rev. Lett.},
  volume = {121},
  issue = {10},
  pages = {107202},
  numpages = {7},
  year = {2018},
  month = {Sep},
  publisher = {American Physical Society},
  doi = {10.1103/PhysRevLett.121.107202},
  url = {https://link.aps.org/doi/10.1103/PhysRevLett.121.107202}
}

@article{Ferrari_PRB_2020,
  title = {{Gapless spin liquid and valence-bond solid in the {${J}_{1}$-${J}_{2}$ Heisenberg} model on the square lattice: Insights from singlet and triplet excitations}},
  author = {Ferrari, Francesco and Becca, Federico},
  journal = {Phys. Rev. B},
  volume = {102},
  issue = {1},
  pages = {014417},
  numpages = {5},
  year = {2020},
  month = {Jul},
  publisher = {American Physical Society},
  doi = {10.1103/PhysRevB.102.014417},
  url = {https://link.aps.org/doi/10.1103/PhysRevB.102.014417}
}

@article{Nomura_PRX_2021,
  title = {{Dirac-Type Nodal Spin Liquid Revealed by Refined Quantum Many-Body Solver Using Neural-Network Wave Function, Correlation Ratio, and Level Spectroscopy}},
  author = {Nomura, Yusuke and Imada, Masatoshi},
  journal = {Phys. Rev. X},
  volume = {11},
  issue = {3},
  pages = {031034},
  numpages = {19},
  year = {2021},
  month = {Aug},
  publisher = {American Physical Society},
  doi = {10.1103/PhysRevX.11.031034},
  url = {https://link.aps.org/doi/10.1103/PhysRevX.11.031034}
}

@article{Wang_ChinPhysLett_2022,
doi = {10.1088/0256-307X/39/7/077502},
url = {https://doi.org/10.1088/0256-307X/39/7/077502},
year = {2022},
month = {jun},
publisher = {Chinese Physical Society and IOP Publishing Ltd},
volume = {39},
number = {7},
pages = {077502},
author = {Wang, Ling and Zhang, Yalei and Sandvik, Anders W.},
title = {{Quantum Spin Liquid Phase in the Shastry-Sutherland Model Detected by an Improved Level Spectroscopic Method}},
journal = {Chin. Phys. Lett.}
}

@article{Liu_PRX_2022,
  title = {{Emergence of Gapless Quantum Spin Liquid from Deconfined Quantum Critical Point}},
  author = {Liu, Wen-Yuan and Hasik, Juraj and Gong, Shou-Shu and Poilblanc, Didier and Chen, Wei-Qiang and Gu, Zheng-Cheng},
  journal = {Phys. Rev. X},
  volume = {12},
  issue = {3},
  pages = {031039},
  numpages = {17},
  year = {2022},
  month = {Sep},
  publisher = {American Physical Society},
  doi = {10.1103/PhysRevX.12.031039},
  url = {https://link.aps.org/doi/10.1103/PhysRevX.12.031039}
}

@article{Liu_PRB_2024,
  title = {{Tensor network study of the spin-$\frac{1}{2}$ square-lattice ${J}_{1}\text{\ensuremath{-}}{J}_{2}\text{\ensuremath{-}}{J}_{3}$ model: Incommensurate spiral order, mixed valence-bond solids, and multicritical points}},
  author = {Liu, Wen-Yuan and Poilblanc, Didier and Gong, Shou-Shu and Chen, Wei-Qiang and Gu, Zheng-Cheng},
  journal = {Phys. Rev. B},
  volume = {109},
  issue = {23},
  pages = {235116},
  numpages = {15},
  year = {2024},
  month = {Jun},
  publisher = {American Physical Society},
  doi = {10.1103/PhysRevB.109.235116},
  url = {https://link.aps.org/doi/10.1103/PhysRevB.109.235116}
}

@article{Viteritti_PRB_2025,
  title = {{Transformer wave function for two dimensional frustrated magnets: Emergence of a spin-liquid phase in the Shastry-Sutherland model}},
  author = {Viteritti, Luciano Loris and Rende, Riccardo and Parola, Alberto and Goldt, Sebastian and Becca, Federico},
  journal = {Phys. Rev. B},
  volume = {111},
  issue = {13},
  pages = {134411},
  numpages = {15},
  year = {2025},
  month = {Apr},
  publisher = {American Physical Society},
  doi = {10.1103/PhysRevB.111.134411},
  url = {https://link.aps.org/doi/10.1103/PhysRevB.111.134411}
}

@article{Lou_PRL_2007,
  title = {{Emergence of U(1) Symmetry in the 3D $XY$ Model with ${Z}_{q}$ Anisotropy}},
  author = {Lou, Jie and Sandvik, Anders W. and Balents, Leon},
  journal = {Phys. Rev. Lett.},
  volume = {99},
  issue = {20},
  pages = {207203},
  numpages = {4},
  year = {2007},
  month = {Nov},
  publisher = {American Physical Society},
  doi = {10.1103/PhysRevLett.99.207203},
  url = {https://link.aps.org/doi/10.1103/PhysRevLett.99.207203}
}

@article{Okubo_PRB_2015,
  title = {{Scaling relation for dangerously irrelevant symmetry-breaking fields}},
  author = {Okubo, Tsuyoshi and Oshikawa, Kosei and Watanabe, Hiroshi and Kawashima, Naoki},
  journal = {Phys. Rev. B},
  volume = {91},
  issue = {17},
  pages = {174417},
  numpages = {4},
  year = {2015},
  month = {May},
  publisher = {American Physical Society},
  doi = {10.1103/PhysRevB.91.174417},
  url = {https://link.aps.org/doi/10.1103/PhysRevB.91.174417}
}

@article{Shao_PRL_2020,
  title = {{Monte Carlo Renormalization Flows in the Space of Relevant and Irrelevant Operators: Application to Three-Dimensional Clock Models}},
  author = {Shao, Hui and Guo, Wenan and Sandvik, Anders W.},
  journal = {Phys. Rev. Lett.},
  volume = {124},
  issue = {8},
  pages = {080602},
  numpages = {6},
  year = {2020},
  month = {Feb},
  publisher = {American Physical Society},
  doi = {10.1103/PhysRevLett.124.080602},
  url = {https://link.aps.org/doi/10.1103/PhysRevLett.124.080602}
}

@article{Harada_PRL_2003,
  title = {{N\'eel and Spin-{Peierls} Ground States of Two-Dimensional $\mathrm{S}\mathrm{U}(N)$ Quantum Antiferromagnets}},
  author = {Harada, Kenji and Kawashima, Naoki and Troyer, Matthias},
  journal = {Phys. Rev. Lett.},
  volume = {90},
  issue = {11},
  pages = {117203},
  numpages = {4},
  year = {2003},
  month = {Mar},
  publisher = {American Physical Society},
  doi = {10.1103/PhysRevLett.90.117203},
  url = {https://link.aps.org/doi/10.1103/PhysRevLett.90.117203}
}

@article{Beach_PRB_2009,
  title = {{$\text{SU}(N)$ Heisenberg model on the square lattice: A {continuous-$N$} quantum Monte Carlo study}},
  author = {Beach, K. S. D. and Alet, Fabien and Mambrini, Matthieu and Capponi, Sylvain},
  journal = {Phys. Rev. B},
  volume = {80},
  issue = {18},
  pages = {184401},
  numpages = {7},
  year = {2009},
  month = {Nov},
  publisher = {American Physical Society},
  doi = {10.1103/PhysRevB.80.184401},
  url = {https://link.aps.org/doi/10.1103/PhysRevB.80.184401}
}

@article{Harada_PRB_2013,
  title = {{Possibility of deconfined criticality in {SU($N$) Heisenberg} models at small $N$}},
  author = {Harada, Kenji and Suzuki, Takafumi and Okubo, Tsuyoshi and Matsuo, Haruhiko and Lou, Jie and Watanabe, Hiroshi and Todo, Synge and Kawashima, Naoki},
  journal = {Phys. Rev. B},
  volume = {88},
  issue = {22},
  pages = {220408},
  numpages = {4},
  year = {2013},
  month = {Dec},
  publisher = {American Physical Society},
  doi = {10.1103/PhysRevB.88.220408},
  url = {https://link.aps.org/doi/10.1103/PhysRevB.88.220408}
}

@article{Sandvik_AIP_2010,
  author  = {A. W. Sandvik},
  title   = {{Computational Studies of Quantum Spin Systems}},
  journal = {AIP Conf. Proc.},
  volume  = {1297},
  number  = {1},
  pages   = {135--338},
  year    = {2010},
  doi     = {10.1063/1.3518900}
}

@article{Sandvik_PRB_1999,
  title = {{Stochastic series expansion method with operator-loop update}},
  author = {Sandvik, Anders W.},
  journal = {Phys. Rev. B},
  volume = {59},
  issue = {22},
  pages = {R14157--R14160},
  numpages = {0},
  year = {1999},
  month = {Jun},
  publisher = {American Physical Society},
  doi = {10.1103/PhysRevB.59.R14157},
  url = {https://link.aps.org/doi/10.1103/PhysRevB.59.R14157}
}

@Article{DEmidio_SciPost_2023,
	title={{Diagnosing weakly first-order phase transitions by coupling to order parameters}},
	author={Jonathan D'Emidio and Alexander A. Eberharter and Andreas M. Läuchli},
	journal={SciPost Phys.},
	volume={15},
	pages={061},
	year={2023},
	publisher={SciPost},
	doi={10.21468/SciPostPhys.15.2.061},
	url={https://scipost.org/10.21468/SciPostPhys.15.2.061},
}

@article{Sen_PRB_2010,
  title        = {{Example of a first-order N{\'e}el to valence-bond-solid transition in two dimensions}},
  author       = {Sen, Arnab and Sandvik, Anders W.},
  journal      = {Phys. Rev. B},
  volume       = {82},
  number       = {17},
  pages        = {174428},
  year         = {2010},
  doi          = {10.1103/PhysRevB.82.174428}
}

@article{Banerjee_PRB_2011,
  title = {{N\'eel to staggered dimer order transition in a generalized honeycomb lattice {Heisenberg} model}},
  author = {Banerjee, Argha and Damle, Kedar and Paramekanti, Arun},
  journal = {Phys. Rev. B},
  volume = {83},
  issue = {13},
  pages = {134419},
  numpages = {8},
  year = {2011},
  month = {Apr},
  publisher = {American Physical Society},
  doi = {10.1103/PhysRevB.83.134419},
  url = {https://link.aps.org/doi/10.1103/PhysRevB.83.134419}
}

@article{Zhou_PRX_2024,
  title = {{SO(5) Deconfined Phase Transition under the Fuzzy-Sphere Microscope: Approximate Conformal Symmetry, Pseudo-Criticality, and Operator Spectrum}},
  author = {Zhou, Zheng and Hu, Liangdong and Zhu, W. and He, Yin-Chen},
  journal = {Phys. Rev. X},
  volume = {14},
  issue = {2},
  pages = {021044},
  numpages = {22},
  year = {2024},
  month = {Jun},
  publisher = {American Physical Society},
  doi = {10.1103/PhysRevX.14.021044},
  url = {https://link.aps.org/doi/10.1103/PhysRevX.14.021044}
}

@article{Ma_PRB_2020,
  title = {{Theory of deconfined pseudocriticality}},
  author = {Ma, Ruochen and Wang, Chong},
  journal = {Phys. Rev. B},
  volume = {102},
  issue = {2},
  pages = {020407},
  numpages = {6},
  year = {2020},
  month = {Jul},
  publisher = {American Physical Society},
  doi = {10.1103/PhysRevB.102.020407},
  url = {https://link.aps.org/doi/10.1103/PhysRevB.102.020407}
}

@article{Nahum_PRB_2020,
  title = {{Note on Wess-Zumino-Witten models and quasiuniversality in $2+1$ dimensions}},
  author = {Nahum, Adam},
  journal = {Phys. Rev. B},
  volume = {102},
  issue = {20},
  pages = {201116},
  numpages = {4},
  year = {2020},
  month = {Nov},
  publisher = {American Physical Society},
  doi = {10.1103/PhysRevB.102.201116},
  url = {https://link.aps.org/doi/10.1103/PhysRevB.102.201116}
}

@book{Georgi,
  author    = {Howard Georgi},
  title     = {Lie Algebras in Particle Physics: From Isospin to Unified Theories},
  edition   = {2},
  series    = {Frontiers in Physics},
  volume    = {54},
  publisher = {Perseus Books},
  address   = {Reading, MA},
  year      = {1999}
}

@article{Binder1981,
  author  = {Kurt Binder},
  title   = {Finite Size Scaling Analysis of Ising Model Block Distribution Functions},
  journal = {Z. Phys. B},
  volume  = {43},
  pages   = {119--140},
  year    = {1981},
  doi     = {10.1007/BF01293604}
}

@article{Sengupta_PRB_2002,
  author    = {Pinaki Sengupta and Anders W. Sandvik and David K. Campbell},
  title     = {{Bond-order-wave phase and quantum phase transitions in the one-dimensional extended Hubbard model}},
  journal   = {Physical Review B},
  volume    = {65},
  number    = {15},
  pages     = {155113},
  year      = {2002},
  doi       = {10.1103/PhysRevB.65.155113}
}

@article{Hukushima_JPhysSoc_1996,
  author  = {Koji Hukushima and Koji Nemoto},
  title   = {{Exchange Monte Carlo Method and Application to Spin Glass Simulations}},
  journal = {J. Phys. Soc. Jpn.},
  volume  = {65},
  number  = {6},
  pages   = {1604--1608},
  year    = {1996},
  doi     = {10.1143/JPSJ.65.1604}
}

@misc{Feuerpfeil2026,
  title   = {{Unifying Dirac Spin Liquids on Square and Shastry-Sutherland Lattices via Fermionic Deconfined Criticality}},
  author  = {Feuerpfeil, Andreas and Shackleton, Leyna and Maity, Atanu and Thomale, Ronny and Sachdev, Subir and Iqbal, Yasir},
  year    = {2026},
  archivePrefix = {arXiv},
  eprint  = {2601.19980},
  primaryClass = {cond-mat.str-el}
}

@misc{Takahashi_2024,
      title={{SO(5) multicriticality in two-dimensional quantum magnets}}, 
      author={Jun Takahashi and Hui Shao and Bowen Zhao and Wenan Guo and Anders W. Sandvik},
      year={2024},
      eprint={2405.06607},
      archivePrefix={arXiv},
      primaryClass={cond-mat.str-el},
      url={https://arxiv.org/abs/2405.06607}
}

@article{
Song_ADV_2025,
author = {Menghan Song  and Jiarui Zhao  and Meng Cheng  and Cenke Xu  and Michael Scherer  and Lukas Janssen  and Zi Yang Meng },
title = {{Evolution of entanglement entropy at SU($N$) deconfined quantum critical points}},
journal = {Science Advances},
volume = {11},
number = {6},
pages = {eadr0634},
year = {2025},
doi = {10.1126/sciadv.adr0634}
}

\appendix

\section*{End Matter}
\subsection*{SU(N) Algebra}

Here we provide some details of our implementation of the SU($N$) algebra. There are many different conventions for these definitions \cite{Read_PRB_1990,Beach_PRB_2009,Kaul_PRL_2012,Kaul_ARCMP_2013,Okubo_PRB_2015} and we here specify those adopted in our work. Many of the basic definitions are from Ref. \cite{Georgi}. 

The $N(N-1)$ off-diagonal generators consist of $N(N-1)/2$ raising operators and $N(N-1)/2$ lowering operators ($\lambda^\pm$), each with a single nonzero entry in the upper or lower triangular part of the matrix, respectively. The remaining diagonal $N-1$ generators are referred to as the Cartan matrices $H_m$. We use the following convention with $m \in \{1,\ldots,N-1\}$
\begin{equation}\label{Eq:Cartan}
    [H_m]_{ij} = \frac{1}{\sqrt{2m(m+1)}}\sum_{k=1}^m\left(\delta_{ik}\delta_{jk}-m\delta_{i,m+1}\delta_{j,m+1}\right)
\end{equation}
As an example, for SU(2) there is only one Cartan element, $\sigma^z/2$. The SU($N$) generators then comprise the set of raising/lowering and Cartan elements, and are normalized by the convention
\begin{equation}\label{Eq:Norm}
    \text{Tr}\left(T^aT^b\right) = \frac{1}{2}\delta_{ab}.
\end{equation}

The operators that appear in the Hamiltonian, the singlet projector $P_{ij}$ and permutation operator $\Pi_{ij}$, can be written in terms of the generators as in Eqs. \ref{Eq:Ops}. For the singlet projector, we can check this form by acting on singlet and adjoint (triplet) states. First, note that using the total ``spin'' operator
\begin{equation}
    T^2_{ij,\text{tot}} = \left(T^a_i + \bar{T}^a_j\right)^2
\end{equation}
and the quadratic Casimir (for the fundamental and conjugate representations)
\begin{equation}\label{Eq:Casimir}
    C_F \equiv T^a_iT^a_i = \bar{T}^a_j\bar{T}^a_j = \frac{N^2-1}{2N}I
\end{equation}
we can write
\begin{equation}
    T^a_i\bar{T}^a_j = \frac{1}{2}\left(T_{ij,\text{tot}}^2 - 2C_F I\right).
\end{equation}
The total spin operator acts on singlet, $\ket{s}$, and adjoint, $\ket{t}$, states as usual
\begin{subequations}\label{Eq:TotalSpin}
\begin{eqnarray}
    T_{ij,\text{tot}}^2\ket{s} &=& 0,  \\
    T_{ij,\text{tot}}^2\ket{t} &=& N\ket{t}.
\end{eqnarray}
\end{subequations}
Note the agreement with SU($2$): $S^2\ket{t} = S(S+1)\ket{t} = 2\ket{t}$. Putting this all together, we show Eq. (\ref{projdef}) acts as a singlet projector
\begin{subequations}
\begin{eqnarray}
    P_{ij}\ket{s} &=& \left(\frac{1}{N^2} + \frac{2}{N}\frac{N^2-1}{2N}\right)\ket{s} = \ket{s},  \\
    P_{ij}\ket{t} &=& \left(\frac{1}{N^2} - \frac{2}{N}\frac{1}{2N}\right)\ket{t} = 0.
\end{eqnarray}
\end{subequations}
The action of the permutation operator is
\begin{equation}
    \Pi_{ij}\ket{\alpha_i\beta_j} = \ket{\beta_i\alpha_j}
\end{equation}
and so it can be written in terms of only delta functions
\begin{equation}
    \left(\Pi_{ij}\right)_{kl} =  \delta_{il}\delta_{jk}.
\end{equation}
Now, we make use of the following identity for generators in the same representation:
\begin{equation}
\left(T^a\right)_{ij}\left(T^a\right)_{kl} = \frac{1}{2}\left[\delta_{il}\delta_{jk} - \frac{1}{N}\delta_{ij}\delta_{kl}\right]
\end{equation}
This identity can be obtained using the normalization of the generators (Eq. \ref{Eq:Norm}). We can immediately identify the first term on the right as the permutation operator above, and the second term is an identity. Inserting these and rearranging gives the definition in Eq. (\ref{permdef}). 

Note again that the additional $1/N$ factor introduced for each permutation operator in Eq. (\ref{Eq:Ham}) is a convention choice to connect to the usual SU(2) Heisenberg exchange for $\Pi_{ij}/2 = 1/4+\mathbf{S}_i\cdot\mathbf{S}_j$, which we then generalize to SU($N$).

The diagonal Cartan generators $H^a$ are used to compute the order parameters $m_s^a$ and $D_\mu^a$ in Eq.~(\ref{Eq:Order}). This form of the generators is a convention we choose, although no results depend on the convention. For SU($2$), the only Cartan element is $\sigma^z/2$. For SU($3$), explicitly, there are two elements:
\begin{equation}
    H_1 =  \frac{1}{2}\begin{pmatrix}
        1 & 0 & 0 \\
        0 & -1 & 0 \\
        0 & 0 & 0
    \end{pmatrix}
\end{equation}
and
\begin{equation}
H_2 = \frac{1}{2\sqrt{3}}\begin{pmatrix}
    1 & 0 & 0 \\
    0 & 1 & 0 \\
    0 & 0 & -2
\end{pmatrix}    
\end{equation}
For higher SU($N$), the general form is Eq. \ref{Eq:Cartan}.

Our order parameters for SU($N$) systems are $N-1$ dimensional vectors, given that there are $N-1$ Cartan matrices. Therefore to define the Binder cumulant such that $U_2 = 0$ in the disordered state (i.e. where $\langle \mathcal{O}^2\rangle = 0$ for $\mathcal{O} = M_s$ or $\psi$), we generalize Binder's result for a scalar random variable \cite{Binder1981} (from a trivial Gaussian integral) 
\begin{equation}
    \langle \mathcal{O}^4 \rangle = 3\langle \mathcal{O}^2 \rangle^2,
\end{equation}
to a vector order parameter based on the $N-1$ Cartan matrices. For an isotropic and zero mean vector, which is satisfied in the disordered phase, a simple calculation gives
\begin{equation}
    \langle \mathcal{O}^4 \rangle = \frac{N+1}{N-1}\langle \mathcal{O}^2 \rangle^2,
\end{equation}
which agrees with the result for $O(n)$ models (see Eq. 77 in Ref. \cite{Sandvik_AIP_2010}), hence our formula for $U_2$ in Eq.~(\ref{Eq:Binder}).

\subsection*{Mean-Field Calculation}

We can carry out a very simple mean-field calculation to compute the expected transition points in the large-$N$ limit. In the limit $Q \rightarrow 0^+$, the ground state of the system should then be the single-color state, which is an eigenstate of the $X$ term. Denote this state as $\ket{x}$. In the opposite limit $Q \rightarrow \infty$, as stated in Large-$N$ proof in the main text, the ground state is the columnar VBS state, call this $\ket{q}$. 

We can compute the energy of each of these states as a function of the coupling $g \equiv X/Q$ and $N$. For $\ket{x}$, it is an eigenstate of the $X$ term. The $Q$ operators each yield a factor of $1/N^2$, and we get
\begin{equation}
    e_x = \frac{\expec{x}{H}{x}}{QL^2} = -\frac{1}{N^2}\left(g+2\right)
\end{equation}
For $\ket{q}$, we choose the state of be one of the four symmetry-broken columnar VBS states. For $\ket{q}$, half the $X$ operators are diagonal on a single columnar covering, and a quarter of the $Q$ operators are aligned with it, so we get
\begin{equation}
    e_q = \frac{\expec{q}{H}{q}}{QL^2} = -\frac{1}{2}\left(\frac{g}{N^2}+1\right)
\end{equation}
We can compute the expected transition as the point $e_q(g,N) = e_x(g,N)$, and we find the mean-field result
\begin{equation}
    g_c = N^2 - 4,
\end{equation}
which should be exact in the limit $N\to \infty$.

We plot this result in Fig.~\ref{fig:Anisotropy}(c) along with the QMC-extracted transition points, and find very good agreement for $N > 2$, as expected for a strongly first-order transition. It is remarkable that the mean-field result agrees very well (within a few percent) with the QMC transition point already for $N=3$. For $N=2$, the mean-field result $q_c=0$ should not be taken seriously because of the very weak first-order transition, i.e., with strong fluctuations not included in the simple mean-field calculation.

\end{document}